\documentstyle[psfig]{mn}

\newif\ifAMStwofonts
\def\simlt{\mathrel{\spose{\lower 3pt\hbox{$\mathchar"218$}}
     \raise 2.0pt\hbox{$\mathchar"13C$}}}
\def\simgt{\mathrel{\spose{\lower 3pt\hbox{$\mathchar"218$}}
     \raise 2.0pt\hbox{$\mathchar"13E$}}}
\def\go{
\mathrel{\raise.3ex\hbox{$>$}\mkern-14mu\lower0.6ex\hbox{$\sim$}}
}
\def\lo{
\mathrel{\raise.3ex\hbox{$<$}\mkern-14mu\lower0.6ex\hbox{$\sim$}}
}

\def\eg{{\it e.g.\ }}
\def\ie{{\it i.e.\ }}
\def\etal{{\it et al.\ }}

\font\syvec=cmbsy10 			%for boldface nabla
\def\bnabla{\hbox{{\syvec\char114}}}	%bold face nabla 
\def\be{\begin{equation}}
\def\ee{\end{equation}}
\def\bea{\begin{eqnarray}}
\def\eea{\end{eqnarray}}

\def\hw2{{\hat W}^2}

\def\eps2{{\epsilon^2}}

\ifoldfss
  \newcommand{\rmn}[1] {{\rm #1}}

  \ifCUPmtlplainloaded \else
    \NewTextAlphabet{textbfit} {cmbxti10} {}
    \NewTextAlphabet{textbfss} {cmssbx10} {}
    \NewMathAlphabet{mathbfit} {cmbxti10} {} % for math mode
    \NewMathAlphabet{mathbfss} {cmssbx10} {} %  "   "    "
  \fi
  \ifAMStwofonts
    \ifCUPmtlplainloaded \else
      \NewSymbolFont{upmath} {eurm10}
      \NewSymbolFont{AMSa} {msam10}
      \NewMathSymbol{\upi}     {0}{upmath}{19}
      \NewMathSymbol{\umu}     {0}{upmath}{16}
      \NewMathSymbol{\upartial}{0}{upmath}{40}
      \NewMathSymbol{\leqslant}{3}{AMSa}{36}
      \NewMathSymbol{\geqslant}{3}{AMSa}{3E}

    \fi
  \fi
\fi % End of OFSS
\ifnfssone
  \newmathalphabet{\mathit}
  \addtoversion{normal}{\mathit}{cmr}{m}{it}
  \addtoversion{bold}{\mathit}{cmr}{bx}{it}
  \newcommand{\rmn}[1] {\mathrm{#1}}

  \newmathalphabet{\mathbfit} % math mode version of \textbfit{..}
  \addtoversion{normal}{\mathbfit}{cmr}{bx}{it}
  \addtoversion{bold}{\mathbfit}{cmr}{bx}{it}
  \newmathalphabet{\mathbfss} % math mode version of \textbfss{..}
  \addtoversion{normal}{\mathbfss}{cmss}{bx}{n}
  \addtoversion{bold}{\mathbfss}{cmss}{bx}{n}
  \ifAMStwofonts
    \ifCUPmtlplainloaded \else
      \UseAMStwoboldmath
      \makeatletter
      \new@mathgroup\upmath@group
      \define@mathgroup\mv@normal\upmath@group{eur}{m}{n}
      \define@mathgroup\mv@bold\upmath@group{eur}{b}{n}
      \edef\UPM{\hexnumber\upmath@group}
      \new@mathgroup\amsa@group
      \define@mathgroup\mv@normal\amsa@group{msa}{m}{n}
      \define@mathgroup\mv@bold\amsa@group{msa}{m}{n}
      \edef\AMSa{\hexnumber\amsa@group}
      \makeatother
      \mathchardef\upi="0\UPM19
      \mathchardef\umu="0\UPM16
      \mathchardef\upartial="0\UPM40
      \mathchardef\leqslant="3\AMSa36
      \mathchardef\geqslant="3\AMSa3E
    \fi
  \fi
\fi % End of NFSS release 1

\ifnfsstwo
  \newcommand{\rmn}[1] {\mathrm{#1}}

  \DeclareMathAlphabet{\mathbfit}{OT1}{cmr}{bx}{it}
  \SetMathAlphabet\mathbfit{bold}{OT1}{cmr}{bx}{it}
  \DeclareMathAlphabet{\mathbfss}{OT1}{cmss}{bx}{n}
  \SetMathAlphabet\mathbfss{bold}{OT1}{cmss}{bx}{n}
  \ifAMStwofonts
    \ifCUPmtlplainloaded \else
      \DeclareSymbolFont{UPM}{U}{eur}{m}{n}
      \SetSymbolFont{UPM}{bold}{U}{eur}{b}{n}
      \DeclareSymbolFont{AMSa}{U}{msa}{m}{n}
      \DeclareMathSymbol{\upi}{0}{UPM}{"19}
      \DeclareMathSymbol{\umu}{0}{UPM}{"16}
      \DeclareMathSymbol{\upartial}{0}{UPM}{"40}
      \DeclareMathSymbol{\leqslant}{3}{AMSa}{"36}
      \DeclareMathSymbol{\geqslant}{3}{AMSa}{"3E}
    \fi
  \fi
\fi % End of NFSS release 2

\ifCUPmtlplainloaded \else
  \ifAMStwofonts \else % If no AMS fonts
    \def\upi{\pi}
    \def\umu{\mu}
    \def\upartial{\partial}
  \fi
\fi

\title[Simulations Using Trees in Fields] 
{Simulations of Spheroidal Systems
with Substructure: Trees in Fields}

\author[Vine and Sigurdsson]
{S.\ Vine \&\ S.\ Sigurdsson\\ 
Institute of Astronomy, Madingley Road, Cambridge CB3 0HA}

\date{Received ** *** 1996; in original form 1996 *** **}

\pagerange{\pageref{firstpage}--\pageref{lastpage}}
\pubyear{1996}

\begin{document}

\maketitle

\label{firstpage}

\begin{abstract}
We present a hybrid technique of N-body simulation to deal
with collisionless stellar systems having an inhomogeneous global
structure.  We combine a treecode and a self-consistent field code
such that each of the codes model a different component of the system
being investigated.  The treecode is suited to treatment of
dynamically cold or clumpy systems which may undergo significant evolution
within a dynamically hot system.  The hot system is appropriately
evolved by the self-consistent field code.  This combined code is
particularly suited to a number of problems in galactic
dynamics. Applications of the code to these problems are briefly
discussed.
\end{abstract}

\begin{keywords}
methods: numerical -- stellar dynamics -- galaxies: kinematics and
dynamics -- galaxies: interactions
\end{keywords}

\section{Introduction}
We concern ourselves with the general dynamical evolution of
a self-gravitating stellar system dominated to the first-order by a stable
or slowly evolving spheriodal mass distribution but with significant
and dynamically distinct substructure. We can refer to this as an
inhomogeneous spheroidal system, \eg a disk or other structure embedded
within a dark matter halo, an encounter between an elliptical galaxy
and less-massive companion, sinking satellites, rings, and fine
structure in ellipticals.  The aim in this work is to develop an
efficient and handy method of modelling the complex dynamical
evolution of the types of systems mentioned, given the resources which
are commonly at hand.

An established method of investigating the dynamical evolution of
stellar systems is through the use of N-body simulations.  Such
simulations have become increasingly sophisticated over the last
decade, enabling detailed experiments to be performed on N-body models
of stellar systems. Improvements have been a result of greater
computing power and more cunning algorithms for following the time
evolution of a system of particles.  A major consideration for the
researcher is to match the computing resources available to an N-body
method suited to modelling the system under consideration.

The majority of numerical treatments of stellar dynamics rely upon the
assumption that stellar systems on the scale of galaxies are collision
free.  This is based on the fact that the two-body relaxation time of
a star, $t_{\rm relax}$, is many magnitudes larger than the age of the
galaxy which contains it.  The appropriate globally averaged estimate
is given by,
\label{relax}
\be t_{\rmn relax} \sim \frac{0.1N}{\ln N} \times t_{\rmn cross}, \ee
for a system of $N$ bodies where
the crossing time,
$t_{\rm cross}$, is the time for a particle to cross the system once
\cite{bt}.  

%Modern computer simulations of galaxies are able to cope
%with sufficient numbers of particles, $N\sim 10^5$, for
%eqn.{(\ref{relax})} to be valid over a global scale.  

The two-body relaxation rate of a body in self-gravitating systems
depends in part on the local density and velocity dispersion.
Relaxation of the orbits of individual particles will occur more
quickly in regions of higher density or lower velocity dispersion.
Other authors present more detailed discussions of relaxation
processes (\eg Farouki \& Salpeter 1994; Huang \etal 1993).  If one is
concentrating on short timescale evolution in a region of fine
substructure in an otherwise dynamically stable or only slowly
evolving system, one may find that the resolution locally is
insufficient to accurately describe the detailed evolution.  This
provides the motivation to combine simulation techniques which will
deal seperately but efficiently with different components of an N-body
system.

\subsection{The N-body Problem}

The dynamical evolution of a system of collisionless particles is
described by the collisionless Boltzmann  equation,
\be \label{cbe}
\frac{\partial f}{\partial t} +\bmath{v} \cdot \bnabla f -\bnabla
\Phi \cdot \frac{\partial f}{\partial\bmath{v}} = 0, \ee
and Poisson's equation,
\be \label{poiss}
\nabla^2 \Phi = 4 \pi G \rho(\bmath{r}),  \ee
where $f$ is the distribution function of the particles, describing
their position $\bmath r$ and velocity $\bmath v$ at a time $t$, and
$\Phi$ is the gravitational potential at $(\bmath{r},t)$ due to all
the particles.  The density of the system is related to $f$ by
\be 
\rho = \int_{}^{} f d^3{\bmath{v}}. \ee
$G$ is the universal gravitational
constant and hereafter is taken to be unity. The
acceleration can then be found from the potential,
\be \label{accn}
\bmath{a}=-\nabla \Phi. \ee

Apart from a number of exact equilibrium solutions of the
collisionless Boltzmann equations, in general this function of seven
independent variables cannot be solved. A feasible way of solving for
the time evolution of equations (\ref{cbe}) and (\ref{poiss}) is to
construct an N-body realisation of the system by sampling the phase
space $(\bmath{r},\bmath{v})$ $N$ times, subject to the probability
density $f(\bmath{r},\bmath{v})$.  The N-body system of particles is
then evolved according to Newton's laws.  It is desirable for the
sample $N$ to be as large as possible to reduce the effects of
statistical noise in the sample, lessening the effects of numerical
two--body relaxation, and increasing the possible spatial resolution.
Memory and processing time of computing resources constrain the value
of $N$ that is achievable, and thus N-body codes generally employ
various approximations to counter the problems raised with smaller
$N$. In general techniques address either one problem or another, and
so choice of appropriate algorithms is very important. One must ensure
that the limitations of any particular algorithm does not invalidate
it's application to the physical system under consideration.

In the following section we outline some of the major techniques that
have been employed in the study of stellar dynamical problems [see
also Hernquist (1987) for a comprehensive review], and explain why it is
we implemented the two methods used in SCFTREE.

\subsection{Techniques}
\label{techs}
Perhaps the most straightforward way to evolve the system is to
calculate directly all the interparticle accelerations. The combined
acceleration, ${\bmath a}_i$, on a particle is
\be \label{force}
{\bmath a_i} = \sum_{j\neq
i}^N\frac{m_j({\bmath{r}_j}-{\bmath{r}_i})}
{[|{\bmath{r}_j}-{\bmath{r}_i}|^2+\epsilon^2]^{3/2}}, \ee
where ${\bmath{r}_i}$, ${\bmath{r}_j}$, $m_i$, and $m_j$ are the
positions and masses of the particles $i$ and $j$, $\epsilon$ is the
softening parameter.  These particle--particle (PP) or direct
summation methods have two important aspects in their favour, the
resolution scale is determined solely by $\epsilon$, and there are no
constraints on the geometry of the system.  The greatest drawback is
in computational cost, at best the CPU time scales as ${\cal
O}(N^2)$. Integration techniques have become quite efficient
\cite{aa72,aa85}, but CPU expensive for $N\go 10^4$ (although new
dedicated hardware has improved the situation, see
below).
  
An extension of the PP technique is the particle--mesh (PM) technique
\cite{he81}. This method imposes a grid upon the system and densities
are assigned to each grid cell.  Applying Fast Fourier Transform
\cite{ct65} to PM methods makes them highly efficient at dealing with
large numbers of particles, thus minimizing statistical fluctuations
in particle distribution.  However spatial resolution is constrained
by the grid spacing.  A hybrid technique has been developed
incorporating PP and PM schemes, unsurprisingly referred to as P$^3$M
\cite{eh74}. This technique has been usefully implemented for
simulations of large-scale structure, where high density contrasts are
expected \cite{ee81}, however when densities get high the number of
close neighbours, which are dealt with by the PP algorithm, becomes
large and prohibitively lengthen the computation time.  Geometric
constraints are also imposed by the existence of the fixed grid.
Further refinement to the P$^3$M technique has been to introduce an
adaptive grid (AP$^3$M, Couchman 1991, Couchman \etal 1995). Examples
of further variants of the adaptive mesh technique are the adaptive
Particle--Multiple--Mesh (PM$^2$) of Gelato, Chernoff, \& Wasserman
(1996), and the hydrodynamic and N-body unstructured adaptive mesh of
Xu (1995).

For systems with a fairly high degree of symmetry it is possible to
represent the potential of the system as a series of terms in a
multipole expansion about the centre of symmetry of the system.
Various basis functions for the expansion have been employed
\cite{cb73,vv77,mg84,ho92}, depending on the global geometry of the
system being modelled.  The expansion is truncated at a specified
order, $n$, this governs the effective resolution of the technique.
The primary advantage of this technique is that computation time
scales as ${\cal O}(nN)$, making large $N$ an attractive
possibility. Weinberg \shortcite{w96} has recently presented an
advance on the expansion technique whereby the expansion basis and
number of expansion terms are matched to the system {\em during} its
time-evolution. The main drawback is that expansion techniques do not
consider individual particle-particle interactions and so local
substructure is ostensively suppressed.  The SCF method of Hernquist
\& Ostriker~(1992) (hereafter HO), is particularly suited to the
observed mass distribution in ellipticals, we expand upon this method
in some more detail in \S{\ref{SCF}}.

The advantages of the particle in a field approach of expansion
techniques and the geometrical flexibility of PP methods are combined
in what are generically described as tree codes.  At close range PP
interaction forces are calculated explicitly, but as the range
increases, particles are grouped together in larger and larger
clusters and their combined potentials are approximated by truncated
expansions.  For each individual particle the force on a particle is
the sum of progressively more distant particle--cluster interactions.
These methods are known as tree methods due to the division of the
particle data into successively smaller and smaller clusters until a
single particle is reached.  The advantage behind the method is that
the time scales involved scale as ${\cal O}(N\log N)$, providing
significant improvements in efficiency over PP methods.  There have
been several methods of constructing the tree structure from particle
data [\eg the AJP method \cite{ap81,je85,po85}, the BH hierarchical
three method \cite{bh86}, see also Hernquist \shortcite{h87}, and
Warren \& Salmon \shortcite{ws92}].  The BH method has proved to be
the most popular, its method of tree construction is more organised
and efficient, although not necessarily as accurate as the AJP
approach.  The hierarchical tree method is described in \S\ref{tree}.

Recent modifications \cite{ws93} to the hierarchical tree process has
given greater control over the accuracy of the force calculations for
the same computational cost of ${\cal O}(N\log N)$.  See also Salmon
\& Warren \shortcite{sw94} for detailed discussion on error analysis
of treecodes.

A potentially promising approach has been the Fast Multipole Method (FMM)
\cite{gr87}.  In principle the FMM is similar to other tree methods
described above, but in calculating the potential at a point, the algorithm
passes from the root node to the leaves, accumulating information from
the cells at the same level according to a range criterion.  This
method has been successful as used in two-dimensional, unclustered
applications, but has failed to be as efficient in three-dimensional
applications \cite{sl91}.

Recent advances in computer architecture have seen a shift to
massively parallel machines. Simply put, the computing load is split
between $N$ separate processors simultaneously, increasing the
computational speed by a factor of $N$.  The method by which it is
split varies depending on the algorithms implemented, some being more
suitable than others, such as the SCF method \cite{hs95}.  Although
intrinsically more tricky to parallelise, tree codes have also been
successfully implemented on parallel machines
\cite{sa91,od94,wa94,d96,dd97}.

An exciting prospect in N-body studies has been the introduction of
special purpose computer chips which are dedicated to the force
calculations. In the simplest case, for example in a PP code, one
would supplement calls to the acceleration subroutine directly by
calls to the chip. Details of the range of ``GRAPE'' and ``HARP''
boards that have been developed can be found in Ebisuzaki \etal
\shortcite{eb93}.  These developments have also been implemented in
codes based on Tree algorithms (\eg Steinmetz 1996).

For the generic problem of a spheroidal system with fine
substructure, the various advantages and disadvantages exhibited by
each of the methods point towards a combination of the SCF method and
BH treecode scheme. These have the advantage of having been widely
tested in a number of astrophysical applications and have been
successfully implemented on commonly available unix workstations.
In \S\ref{principle} we outline in more detail the principles behind
the specific Tree and SCF methods used here. Following this we
describe how these two separate algorithms were combined
to form the new hybrid technique, and how we implemented this new {\tt
SCFTREE} code. We thoroughly test
the efficiency and accuracy of {\tt SCFTREE}, and present the results
in \S\ref{tests}. Finally in \S\ref{discussion} we discuss the
range of applicability of this new technique to a wide variety of
dynamical problems.

\section{Putting Trees in Fields}
\label{principle}
\subsection{The Hierarchical Tree Code}
\label{tree}
%{\bf Summarise stuff from \cite{h87} which reviews tree codes very well.}
%{\bf Possibly include something along the lines of \S(IIc)} 

The nature of the hierarchical tree method \cite{bh86}, as introduced
in the previous section lends itself to  efficient and logical
coding of its algorithm.  

The basic principle in improving the direct summation (PP) technique
is to group together long range interactions.  In the BH tree code
particles are contained within a cubic volume, known as the root
cell. The tree code algorithm gets its name from the way the particles
are grouped together in a hierarchical level of cells, with the root
cell being at the top.  This root cell is subdivided into eight
further cells, the next level down in the hierarchy of cells.  If any
of these cells contain more than one particle, that cell is further
subdivided into eight.  This `tree building' process continues until
the subdivision of cells at a particular level in the tree building
can go no further, \ie the final hierarchical tree level is composed
of cells which contain only one particle.  For all cells on every
level which contain more than one particle, a multipole expansion is
performed about its centre of mass.  This expansion is typically
truncated at the quadrupole term, although some implementations of the
tree algorithm include higher terms, \eg the octupole term \cite{ma93}.
Here we truncate the expansion at the quadrupole term, so that for
each cell we calculate its centre of mass and its quadrupole moment,
{\bf Q}, \cite{g80}.  The acceleration at a position $\bmath{r_c}$
from the centre of mass of the cell can then be expressed as
\be 
\label{quad}
{\bmath a_i} = G\left[-\frac{M_c{\bmath r_c}}{r_c^3} +
\frac{{\bf Q} \cdot {\bmath r_c}}{r_c^5} -
\frac{5}{2}
\frac{({\bmath r_c} \cdot {\bf Q} \cdot {\bmath r_c}) {\bmath r_c}}{r_c^7}
\right]. \ee
For any particle in the system it, in effect, now sees a hierarchy of
cells of different sizes and different distances.  If $s$ is the size
of a cell and $d$ is the distance from the particle to the cell, the
simple criterion,
\be
\label{crit}
\theta > \frac{s}{d}, \ee 
where $\theta$ is known as the tolerance parameter, can be applied to
decide whether the interaction between particles in that cell and
the one particle is adequately approximated by the acceleration given
by equation(\ref{quad}), or whether the particle should `look' at the
lower level of smaller cells.  If the cell contains a single
particle then the acceleration is simply given by equation
(\ref{force}). The tolerance parameter, $\theta$, can be set by the
investigator to control the accuracy of the approximation.  The usual
process of calculating forces on a particle is to `walk' through the
tree from the root cell downwards forming a list of valid interactions
between cells and the particle.  The length of this list is
proportional to $\log N$, hence the CPU time for a single timestep
scales with ${\cal O}(N\log N)$.  This is what makes the tree method a
distinct improvement over the PP method.

\begin{figure*}
\psfig{figure= 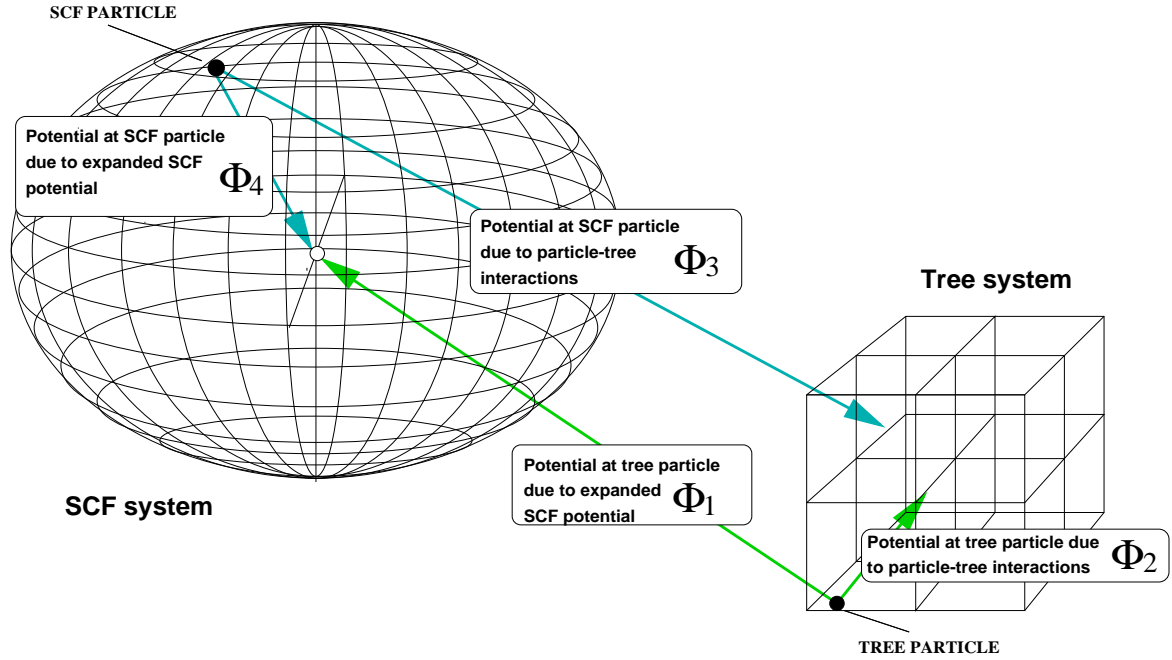,width=6.8in}
\caption{Schematic representation of the interaction between the {SCF}
system and {Tree} system.  The potentials $\Phi_1$ and $\Phi_4$ are
due to the expanded SCF potential, see equation (\ref{scfpot}), at a
position relative to the centre of mass of the SCF system (as defined
in \S\ref{scftree}).  The potentials $\Phi_2$ and $\Phi_3$ are due to
the tree system of particles. The calculation of the tree potential is
described in \S\ref{tree}.  }
\label{scheme}
\end{figure*}

\subsection{Self-Consistent Field Code}
\label{SCF}
%Discuss SCF ad nauseum. \cite{cb73,vv77,ho92,sh95,hh95,hs95} {\bf
%\noindent Insert here \S2 and \S2.1 from \cite{hs95}
%} Describes the principle and details of the SCF method.

%Sigurdsson, He, Melhem and Hernquist, Sigurdsson and Mihos.
%Also variants: Earn, Syer, Saha, Weinberg.

As far as a star in a galaxy is concerned, the dominant component of
the gravitational potential which it experiences comes from the
averaged field of billions of distant gravitational sources.  It may
appear rather odd that on the whole, the simulation of the
gravitational dynamics of galaxies employs a method of basically
summing up all the individual particle--particle interactions, each
one of which is just about negligible.  The expansion techniques
described in \S\ref{techs} appear more natural, as the force
calculation on each particle considers the effect of the {\em mean}
gravitational field of the whole system directly acting on the
particle.  One might expect that the effects of two-body relaxation
would be ostensively mitigated.  On the other hand the model system
still consists of many fewer bodies than a real system, and thus the
statistical fluctuations brought about by the finite number of
particles will cause the expanded gravitational field to fluctuate and
thus the overall effect on the system approaches that of one that has
experienced a corresponding amount of two-body relaxation.  The major
advantage of the SCF method is that the speed of the computation
scales as ${\cal O}(N)$.  Therefore it is possible to increase the
practical limit on the number of particles, $N$, and thus any
statistical fluctuations in the expanded potential are reduced.

The principle behind the SCF approach is to represent the potential of
the system as a truncated series of terms of an expanded set of {\em
basis} functions.  It is possible to find the coefficients of such an
expansion from the known density field `sampled' by the particles.
Poisson's equation is solved for the set of basis functions, and
coefficients of the expanded potential can be found.  From the
potential expansion the acceleration of any particle can be directly
calculated.  This technique had been applied in various guises and
various degrees to a limited extent in the past \cite{cb73,vv77,mg84}.
The philosophy behind the SCF method of Hernquist \& Ostriker~(1992)
 was to use a set of basis function where the lowest
order terms well represent the observed distribution in spheroidal
galaxies.  This method has subsequently been successfully utilised in
further studies, [\eg Johnston, Spergel, \& Hernquist (1995); Hozumi
\& Hernquist \shortcite{hh95}, Hernquist, Sigurdsson, and Bryan
\shortcite{hs95}, Sigurdsson, Hernquist, \& Quinlan \shortcite{sh95}].
Below we present a quick resum\'e of the method of \cite{ho92}.

Hernquist \shortcite{h90} demonstrated that a density-potential pair
exists such that their properties provide a very good approximation to
the actual properties of observed galaxies, \ie the $R^{1/4}$ law.
They are:
\be \label{hrho}
\rho(r)=\frac{M}{2\pi}\frac{a}{r}\frac{1}{(r+a)^3}; \ee
and
\be \label{hphi}
\phi(r)=-\frac{M}{r+a}. \ee
The total mass is $M$, and $a$ is the scale length related to the
half-mass radius such that $r_{1/2}=(1+\sqrt{2})a$.  A great advantage
of this density--potential pair is that many of its properties can be
derived analytically.

A biorthogonal {\em basis set} is constructed for the density and
the potential, and they can be written as expanded series:
\be \rho({\bmath r}) = \sum_{nlm} A_{nlm} \rho_{nlm}({\bmath r}), \ee 
and 
\be \label{scfpot}
\Phi ({\bmath r}) = \sum_{nlm} A_{nlm} \Phi_{nlm}({\bmath r}), 
\ee 
where $n$, $l$, and $m$ are equivalent to ``quantum'' numbers in the
expansion, one radial, and two angular.  Each pair $\rho_{nlm}$ and
$\Phi_{nlm}$ satisfy Possion's equation (\ref{poiss}).

The lowest order terms are set to be those of the assumed model, \ie
$\rho_{000}\equiv\frac{1}{2\pi}\frac{1}{r}\frac{1}{(1+r)^3}$ and
$\Phi_{000}\equiv-\frac{1}{1+r}$, where $G=M=a=1$.  It is possible to
construct the higher order terms of the expanded potential-density
series $\Phi_{nlm}({\bmath r})$ and $\rho_{nlm}({\bmath r})$, and to
derive an expression for calculation of the coefficients $A_{nlm}$.
The interested reader may refer to the clear presentation given in
Hernquist \& Ostriker \shortcite{ho92}.

Once all the $A_{nlm}$ have been calculated from the known coordinates
of all the particles, the potential at the location of any particle
can be evaluated using equation (\ref{scfpot}), and thus the
acceleration of the particle can be found using equation (\ref{accn}).
%%
%\be \label{scfpot2}
%\Phi (r_k, \theta _k, \phi_k) = \sum
%_{nlm} A_{nlm} \Phi _{nl} (r_k) \, Y _{lm} (\theta _k, \phi _k) \, .
%\ee
% 
%The components of the acceleration are then be obtained by simple
%analytical differentiation of equation (\ref{scfpot2}).

For the purposes of our hybrid code, the basis set derived by
Hernquist \& Ostriker (1992) is an ideal choice.  Other basis sets may
be more useful for other systems such as discs, or objects that are
not well-approximated by the $R^{1/4}$ profile.  [Other basis sets are
discussed elsewhere \eg Saha (1993); Earn (1995); Earn \& Sellwood
(1995); Zhao (1996); Syer (1995)]

\subsection{Combining Tree and SCF}
\label{scftree}
The problem we are faced with is how to combine the two codes. It is a
relatively simple task, all we need to achieve is 
\be \label{atree}
{\bmath a}_{\rmn (on\,tree\,particle)} = {\bmath a}_{\rmn (due\,to\,Tree)} + 
{\bmath a}_{\rmn (due\,to\,SCF\,expansion)}, \ee 
\be \label{ascf}
{\bmath a}_{\rmn (on\,SCF\,particle)} = {\bmath
a}_{\rmn (due\,to\,SCF\,expansion)} + {\bmath a}_{\rmn (due\,to\,Tree)}.  \ee 
The left-hand sides of equations (\ref{atree}) and (\ref{ascf}) are
the total accelerations on a particle in the TREE and SCF systems
respectively.  The first terms on the right hand side are the
accelerations on the particles due to their own systems, the terms on
the far right are the extra accelerations due to the other system of
particles.  This is schematically shown in Figure (\ref{scheme}). 

When implementing this there are a couple of points to bear in mind.
Firstly we need to take account of the response of the SCF system as a
whole to the TREE system, \ie the centre of mass will not be fixed.
So we have to make sure that the SCF expansion is calculated about the
centre of mass, or symmetry, of the SCF system of particles.  As the
SCF system evolves it may become asymmetric to some degree, hence we
need to ensure that the expansion takes place about the most tightly
bound region.  We achieve this by calculating the centre of mass only
from the contribution of particles which have less than the average
particle energy.  This is a parameter which may affect the accuracy of
the expansion.  It is possible to vary the energy level cut off for
calculation of centre of mass.  We discuss this further in
\S\ref{momenta}.

Our concern now is how to treat the Tree system of particles with
respect to the individual SCF particles.  There are two options.
Firstly, we may perform an expansion of the tree system much in the
same way as the SCF system, then each SCF particle will view the
potential due to the Tree system as a truncated set of terms in an
expanded potential.  For encounters that are non-penetrating this may
prove a quick and reasonable approximation, but because the tree
system is {\em a priori} not assumed to have any particular geometry,
and likely to have a widely disrupted one, we would not expect it to
be accurately represented in just a few terms of an expanded
potential.  Since we also desire our models to describe encounters and
interacting systems we have to be aware that particle-particle
interactions would not be accounted for if we used an SCF expanded
Tree potential, \eg the effects of dynamical friction would be poorly
represented.  Our other, more instinctive, option is to treat the
system of Tree particles simply as a tree from the point of view of
the SCF particles.  This is a much more logical procedure as the tree
has already been constructed and all that needs to be done for each
SCF particle is to construct a tree interaction list and then sum the
over particle-cell forces in the list.

The main calculations performed by the code in each time
step of the system are outlined in the following sections.

\subsubsection{Expansion Centre of the SCF System}
As stated above the centre of mass of the SCF system is updated every
timestep. Its position is calculated from the most tightly bound
particles and is subsequently used as the origin of the SCF expansion
when producing the coefficients, $A_{nlm}$.

\subsubsection{Tree Particle -- SCF Interaction}
\label{tscf}
The position of each tree particle, ${\bmath
r}\rightarrow(r,\theta,\phi)$, with respect to the SCF expansion
centre, is passed to a routine which calculates $\Phi({\bmath r})$,
\ie equation (\ref{scfpot}). The $A_{nlm}$ have been already found for
the case of the SCF system.  The resulting acceleration of the Tree
particle due to the SCF system is subsequently calculated.

\subsubsection{Acceleration of Tree Particles due to Tree}
As described in \S\ref{tree}, a tree data structure is built from the
particles in the tree system.  Then for each particle an list of
interactions between itself and the tree cells is formed subject to
the tolerance parameter, $\theta$.  Finally the cell-particle
accelerations for each element in the list are calculated, summed, and
added to the SCF acceleration from \S\ref{tscf}.

\subsubsection{Acceleration of SCF Particles due to SCF}
As described in \S\ref{SCF} the coefficients, $A_{nlm}$, of the
SCF expansion are calculated from the positions of the SCF particles.
The acceleration due to the SCF system on each SCF particle is then found by
applications of equations (\ref{scfpot}) and (\ref{accn}).

\subsubsection{SCF Particle -- Tree Interaction}
For each SCF particle we build a tree interaction list. Each
level of tree cells from the largest (root) to the smallest (single
particle) is examined.  If the criterion of inequality (\ref{crit}) is
satisfied then the cell is added to the interaction list, otherwise
the next level down is examined.  Once the interaction list is formed,
the list is looped through, summing the accelerations between the SCF
particle and the tree cell as calculated by equation (\ref{quad}),
and adding this to the acceleration due to the SCF particles
themselves.

\subsubsection{summary}

\noindent For each timestep:
\begin{enumerate}
\renewcommand{\theenumi}{(\arabic{enumi})}
\item Find centre of mass of SCF system to use as centre\\ 
\hbox{\hspace{1cm}}of expansion.
\item Calculate the coefficients $A_{nlm}$ of the terms in the\\
\hbox{\hspace{1cm}} SCF expansion.
\item Build the Tree from all the Tree particles\\
\item For each Tree particle:\\
\hbox{\hspace{1cm}} a) Calculate $\Phi_{SCF}(\bmath{r}_{Tree})$.\\
\hbox{\hspace{1cm}} b) Form Tree interaction list between TREE particle\\
\hbox{\hspace{1.5cm}}    and Tree system subject to $\theta$.\\
\hbox{\hspace{1cm}} c) Calculate $\Phi_{Tree}(\bmath{r}_{Tree})$.\\
\hbox{\hspace{1cm}} d) Calculate acceleration of Tree particle.
\item For each SCF particle:\\
\hbox{\hspace{1cm}} a) Calculate $\Phi_{SCF}(\bmath{r}_{SCF})$.\\
\hbox{\hspace{1cm}} b) Form Tree interaction list between SCF particle\\
\hbox{\hspace{1.5cm}} and Tree subject to $\theta$.\\
\hbox{\hspace{1cm}} c) Calculate $\Phi_{Tree}(\bmath{r}_{SCF})$.\\
\hbox{\hspace{1cm}} d) Calculate acceleration of SCF particle.
\item Update positions and velocities of all particles.
\end{enumerate}

\subsection{Implementation}
The aim in this project was to combine two separate codes into one
hybrid code.  The process was made much easier by the availability to
us of optimised versions of both the treecode and SCF code written in
Fortran 77.  We acknowledge the author, Lars Hernquist, for making
these codes freely available.
 
The two codes were made compatible and carefully purged of any
overlapping routines and variables ensuring no cause for confusion
existed. The extra routines added were: calculation of the SCF
system centre of mass correction; formation of tree interaction list
for an SCF particle; calculation of the acceleration on a SCF particle
from the tree list; calculation of the acceleration on a Tree particle
from the SCF expanded potential; and a number of modifications to the
output information. 

The final version of the SCFTREE code was implemented on a Sun Sparc20
workstation running Solaris 2.4, and being written in standard F77
should be fully portable.

\section{Tests}
\label{tests}
Having two sets of particles that are evolved under two different
numerical schemes one has to be very careful that the system as a
whole is behaving as a realistic model of the system it represents.
We conduct a variety of stringent tests on accuracy and efficiency
which display the applicability of this code to the systems it was
designed to deal with.  In this section we quantify the efficiency of
the code, check its validity with respect to the constants of motion,
and put its performance to the test in a number of dynamical
examples.

\subsection{Models}
\label{models}
Three distributions of particles are realised in order to test the
stability of spherical systems and compare the ability of SCFTREE to
handle distributions departing from the Hernquist distribution (the
SCF basis set).  The distributions are the Hernquist density profile
\cite{h90}, the Plummer density profile \cite{pl11}, and the Lowered
Evans model \cite{e93}.  The Lowered Evans model has been shown to be
useful for the practical modelling of dark halos \cite{kd94}, and as
such is our preferred model in many of the tests.  With reference to
the parameters in Kuijken \& Dubinski (1994) we use the following for
all our Lowered Evans models: $\Psi_0=-5.0$, $v_0=1.5$, $q=1.0$,
$(r_c/r_k)^2=1.0$, $r_a=1.0$.

In these tests the models are populated with up to $10^5$ particles
and a specified fraction of them are designated as particles to be
dealt with by the Tree part of the code, referred to hereafter as
`Tree particles'.  The remainder of the particles are dealt with by
the SCF part of the code and referred to as `SCF particles'.

Analogous to the investigations of Hozumi and Hernquist
\shortcite{hh95} who perform tests on the pure SCF code, we examine
the behaviour of SCFTREE in systems which are not in equilibrium.  For
this purpose we construct a uniform density sphere with velocity
dispersions assigned according to a specified initial virial ratio,
the evolution of the system is followed until it reaches its final
relaxed state.

Extending these tests of SCFTREE to more interesting systems we
perform test runs on a disk system.  A disk, bulge, and halo model is
set up as prescribed in Dubinski and Kuijken (1995). Such a system is
well suited to the SCFTREE technique whereby the disk particles are
assigned to the Tree and the halo and bulge particles are assigned to
the SCF system. Here we merely examine the effect of increasing the
number of halo particles on the stability of the disk.

%For the following checks on timing of the code, and tests of conserved
%quantities, we constructed non-rotating equilibrium models as
%prescribed in Kuijken \& Dubinski \shortcite{kd94}.  These are
%referred to as lowered Evan's models.  In this case the models were
%set up to be as spherical as possible. The timing models were
%variously populated with up to 100,000 particles.
%
%For the collapse tests, \S\ref{collapse} the initial model was a
%uniform spherical distribution of particles with random velocities,
%normalised to the kinetic energy of a virialised system of the same
%mass.
%
%For the dynamical tests we constructed models based on the method given in
%Dubinski \& Kuijken \shortcite{dk95}.

%\begin{table*}
%\begin{center}
%\caption{Details of the various initial models used in the tests.}
%\label{testpars}
%\begin{tabular}{r@{ = }l | r@{ = }l | r@{ = }l | r@{ = }l }
%\hline
%\multicolumn{2}{c|}{Model A} & \multicolumn{2}{c|}{Model B} &
%\multicolumn{2}{c|}{Model C} & \multicolumn{2}{c}{Model D} \\
%\hline \hline
%$\theta$ & $0.7$ \\
%$dt$ & 0.05 \\
%$\epsilon$ & 0.1 \\
%$n_{steps}$ & 1000 \\
%$l_{max}$ & 4 \\
%$n_{max}$ & 6 \\ \hline
%DF & Lowered Evans & Hernquist & potential\\
%$N$ & 20000 \\
%Tree & 10\% \\ 
%
%\hline
%\end{tabular}
%\end{center}
%\end{table*}

\begin{figure*}
\label{timingfig}
\psfig{figure= 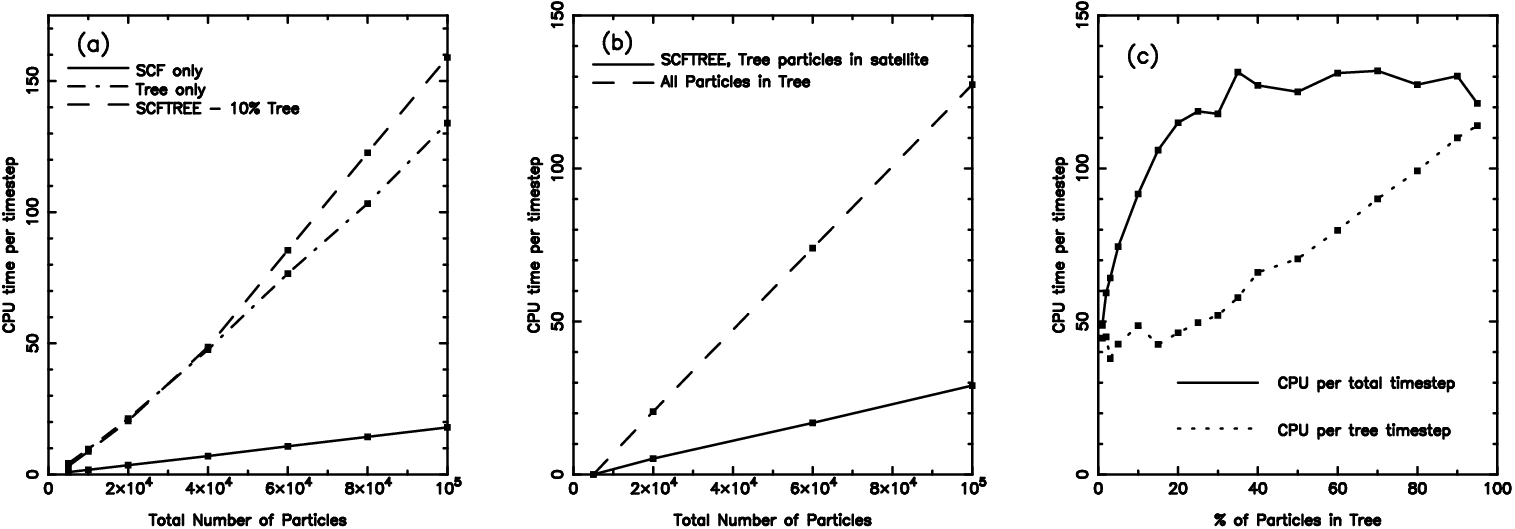,width=17.5cm,angle=270}
\caption
{CPU per timestep performances for SCFTREE.  \protect\newline
(a) SCFTREE is compared to pure
Tree and pure SCF codes, in the SCFTREE runs 10\% of the particles are
designated as Tree particles and distributed randomly through the
system (A lowered Evans model with 20000 particles, dt=0.05). \protect\newline 
(b) Binary system of spherical galaxies in a circular orbit. Mass
ratio 10:1.  In the SCFTREE runs the smaller satellite is composed of
only Tree particles and the larger body composed of only SCF
Particles. \protect\newline 
(c) Variation of the fraction of Tree particles, $N_{\rm
tree}/N_{SCF}$ distributed randomly in a Lowered Evans
model. \protect\newline}
\end{figure*}

\subsection{Timing}
The individual timing performances of tree and SCF codes have been
shown to scale to ${\cal O}(N_{tree}\log N_{tree})$, \cite{bh86}, and
${\cal O}(N_{SCF})$, \cite{ho92}, respectively.  For the combined code
we have to consider the CPU time spent on the interaction between the
two components.  The calculation of the Tree force on each SCF
particle would, at worst, scale as ${\cal O}(N_{SCF}\log N_{tree})$,
and that of the SCF force on the tree particles  scales as ${\cal
O}(N_{tree})$.  So overall we would expect the CPU time to scale as
\be
\label{cpu_scale}
{\cal O}\left[ N_{SCF}(1+\log N_{tree}) + N_{tree}(1+\log N_{SCF})\right].
\ee
As the major advantage of SCFTREE is expected to be its efficiency in
dealing with systems with substructure, we perform tests varying the
fraction of tree particles in the system and changing the initial
distribution of these particles.  The models used here for the
performance checks are the Lowered Evans models.

We examine how the timing of SCFTREE behaves with the total number of
particles. The fraction of tree particles is set at 10\% of the total
and they are distributed randomly throughout the system.  The random
distribution of particles implies that there will not be any advantage
in time saved due to the Tree structure.  This is demonstrated in
Figure 2(a) where we compare purely Tree, purely SCF, and SCFTREE
timings as the total number of particles is increased up to $10^5$. As
expected no advantage is offered by SCFTREE with randomly distributed
particles over the pure Tree treatment.

The great advantage of SCFTREE is demonstrated in Figure 2(b) where
the Tree particles are assigned to a satellite system orbiting a more
massive system composed of the SCF particles.  This is compared to a
pure Tree treatment of the same binary system.  The Figure clearly
demonstrates the strength of SCFTREE when the Tree particles are
assigned to a distinct substructure.  We see that for SCFTREE the CPU
time per timestep scales approximately as ${\cal O}(N)$, a substantial
improvement over the pure Tree timings.

Finally we examine the performance of the code as the fraction of Tree
particles is increased.  Figure 2(c) shows the CPU time spent both on
the whole SCFTREE timestep and just on the Tree timestep in a system
with a randomly distributed fraction of Tree particles.  It can be
seen that for such a system the fraction of Tree particles must remain
below $\sim 10$\% to gain any reasonable advantages.

%Figure (\ref{timingfig}a) shows how the CPU per timestep scale with
%$N$. ****The dotted line shows the expected performance given in
%equation (\ref{cpu_scale}) scaled with $N=100000$****.  Figure
%(\ref{timingfig}b) shows how the CPU timestep varies as we increase the
%fraction of tree particles in a standard model.  We see that using
%$N_{tree}<15\%$ is desirable.

\subsection{Relaxation}
\begin{figure}
\psfig{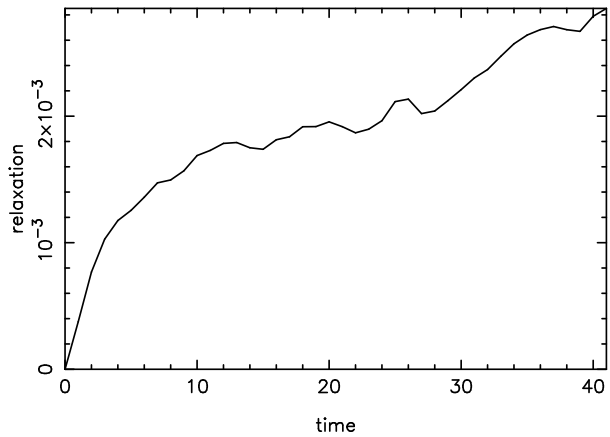}
\caption{The magnitude of 2-body relaxation in the system versus
time. The relaxation measure we use is $\left< |\Delta E/E|^2\right>$,
defined in equation (\ref{eqrelax}). The model is a Lowered Evans
model containing 20,000 particles, 10\% being randomly distributed
tree particles.}
\label{rel}
\end{figure}

\begin{figure}
\centering
\psfig{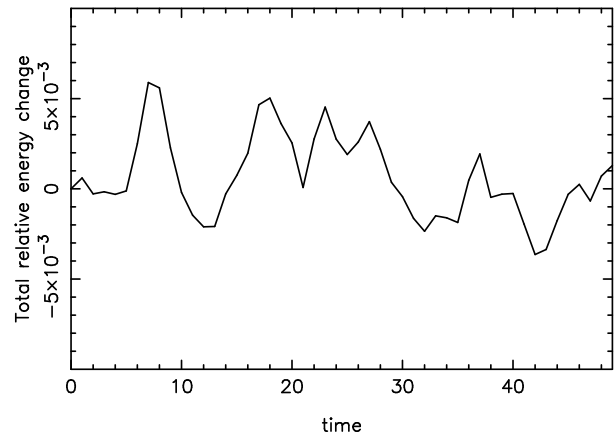}
\caption{An example of the total fractional energy change,
$(E_t-E_0)/E_0$, as a function of time, for a Lowered Evans model,
20000 particles with 10\& in Tree.  SCF truncation parameters are
$n=16$ and $l=6$.  SCFTREE conserves energy to within %0.5$\% for the
duration of this run.}
\label{energy}
\end{figure}

\begin{figure*}
\begin{center}
\psfig{figure= 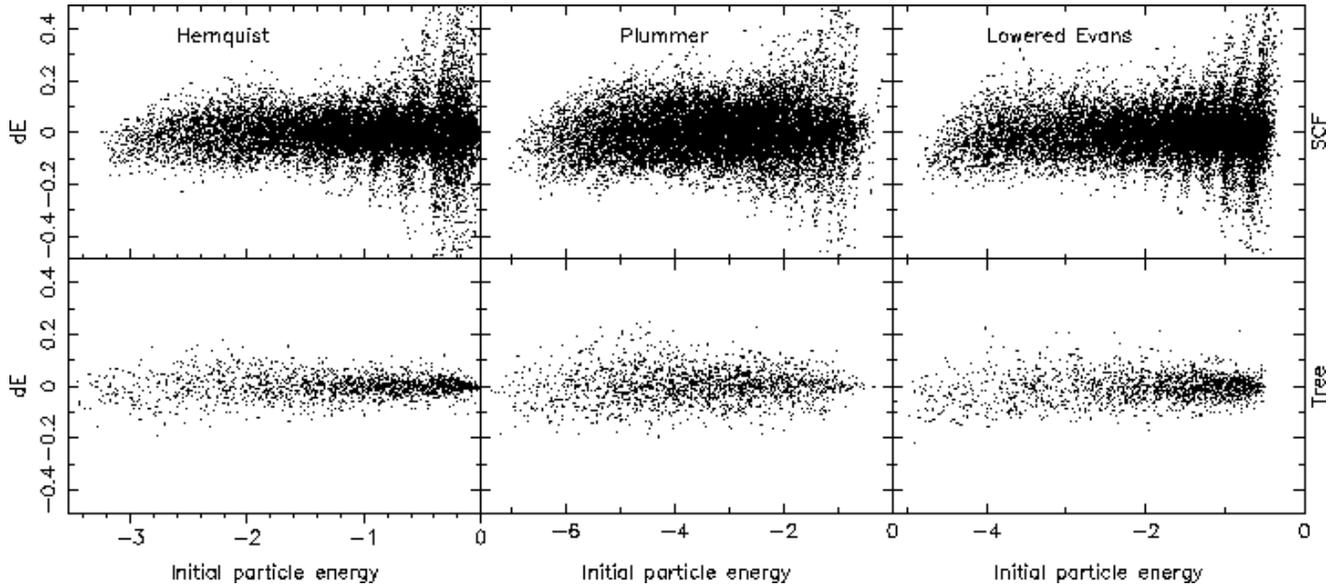,width=17.5cm,angle=270}
\caption{How the individual particle energies behave with respect to
different initial equilibrium models.  Results are calculated from
$t=10$ to $t=50$.  All models consist of a total of 20000 particles,
2000 of which are randomly assigned to the Tree system. The SCF
expansion is truncated at $n=6$ and $l=2$.}
\label{pen}
\end{center}
\end{figure*}

\begin{figure*}
\begin{center}
\psfig{figure= 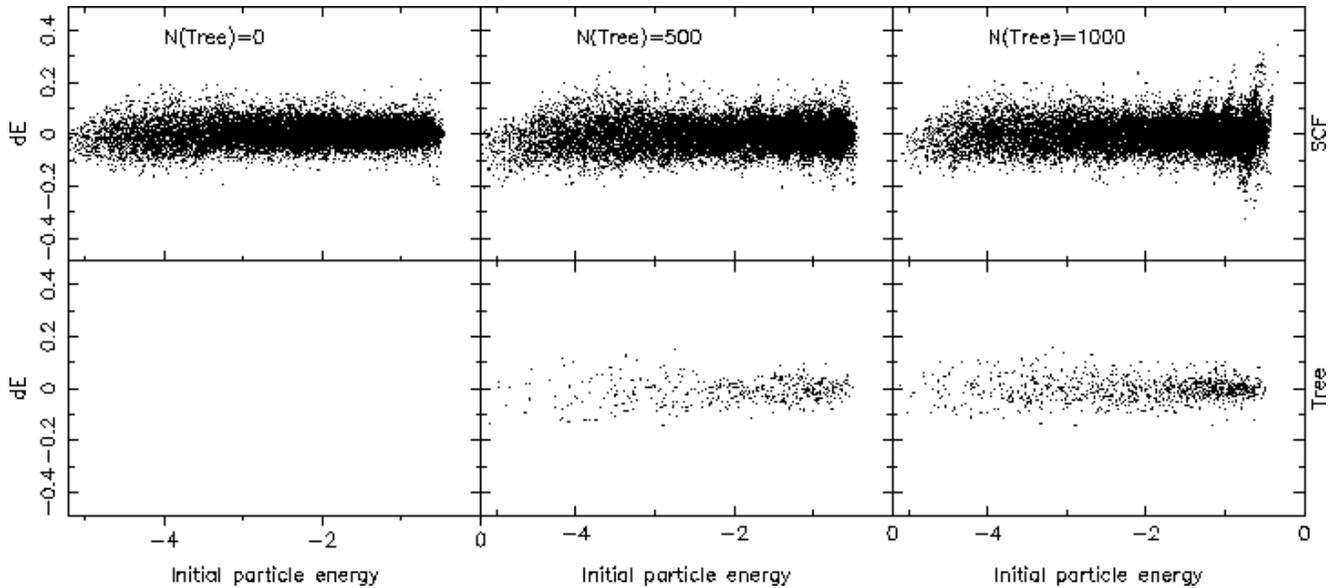,width=17.5cm,angle=270}
\caption{Showing the effect on individual particle errors as the
fraction of Tree particles in the simulation is reduced.  This
demonstrates that it is the presence of the Tree particles in the
system which cause the vertical ``bands'' associated with certain
energies.  The relative energy difference per particle, $dE$, is
calculated from $t=10$ to $t=50$. These models are Lowered Evans
models with a total of 20000 particles.}
\label{pen2}
\end{center}
\end{figure*}

It is instructive to follow the behaviour of two-body relaxation as
the system evolves.  As an indicator of two-body relaxation we take
the mean square fractional energy change of the particles,
$R_{2-body}=\left< | \Delta E/E|^2\right>$. We define this as
\be \label{eqrelax} 
R_{2-body}=\frac{1}{N}\sum_{i=1}^{N}\left| \frac{E_{i,t}-E_{i,0}}{E_{i,0}} 
\right|^2,
\ee
$E_{i,t}$ and $E_{i,0}$ are the total energy (kinetic plus potential)
of particle $i$ at times $t$ and 0 respectively.  In Figure
(\ref{rel}) we see that by our measure the average relaxation
remains less than $1\%$ for the period of the simulation.  The initial
sharp rise is expected due to the initial random placing of the
particles, meaning that many pairs of particles will have initial
separations less than the tree softening length, $\epsilon$.

%\subsection{Constants of Motion}
\subsection{Energy Conservation}
Relaxation tells us how close our models approach the behaviour of a
collisionless system, providing us with a quantitive estimate of over
what timescales our simulations can be applied.

We may also look at the change in the energy of each particle over a
suitable time period and compare the behaviour of SCF and tree
particles. This gives another gauge of relaxation of particles this
time as a function of the binding energy, $E$.  We follow a similar
analysis as HO92 and plot $\Delta E/E$ vs $E$ for both SCF and tree
particles.  Figure(\ref{pen}) shows the results for the 3 spheroidal
models: Hernquist; Plummer; and Lowered Evans.  Each model is
populated with 20000 particles and the energy change is fractional
difference of the particle energy at times $t$ and $t_0$.  One will
notice all three SCF particle plots exhibit a vertical band structure
at certain binding energies.  The cause of some particles being more
liable to undergo two-body relaxation than other is made clearer by
Figure(\ref{pen2}) which shows the same plots for three Lowered Evans
models with 0, 500, and 1000 tree particles respectively.  The greater
the fraction of tree particles the wider and more distinct the
vertical features that appear in the regions of low binding energies.
The first point to bear in mind is that at low energies the fractional
energy changes will become large because the binding energy is
approaching zero.  Secondly it is clear that greater the number of
tree particles the greater the fractional energy change.  This is due
to the fact that individual SCF particles interact directly with the
tree particles, and so the more tree particles the greater the
energy exchange between the two components.  Finally the bands that are
seen appear as a result of truncating the SCF expansion.  The expanded
potential of the system is derived from the actual positions of all
the constituent SCF particles.  The actual binding energy of a
particle may not agree exactly with the potential given by the
expansion, causing an error in the resulting acceleration.  There will
be certain values of binding energy which correspond to the
divergences in the truncated expanded potential from the true value,
thus the particles at those binding energies will experience greater
errors in acceleration.  Clearly the presence of Tree particles in the
system exacerbates this effect.  However because these effects are
symmetric the overall global errors remain minimal.

This is confirmed by monitoring the total energy of the system.
Figure(\ref{energy}) traces the fractional variation in total energy
of the system, $(E_t-E_{t_0})/E_{t_0}$.  The model is a Lowered Evans
with 20000 particles, 10\% being Tree particles, and timestep
$dt=0.05$.  Energy is conserved in this system to within 0.8\% over a
period of 50 time units.

\subsection{Momenta}
\label{momenta}
\begin{figure}
\psfig{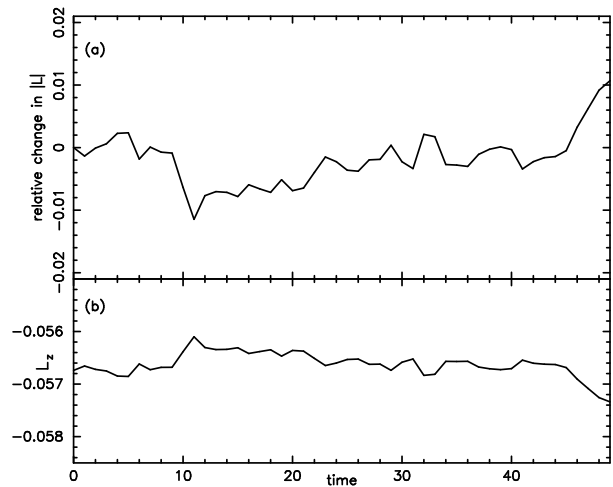}
\caption{Plot (a) shows the relative angular momentum change,
$(|L_t|-|L_0|)/|L_0|$, with time.  Plot (b) shows the evolution of the
z-component of the angular momentum, $L_z$, with time.  The data is
from a Lowered Evans model containing 20,000 particles, 10\% of which
are tree particles.}
\label{angmom}
\end{figure}
As with all expansion codes angular and linear momentum are
intrinsically not conserved exactly, due to approximations in the force
calculation.  We might expect it to be more pronounced in the SCFTREE
case where the particles in the Tree and SCF codes do not respond to
each other equally and oppositely.  Linear momentum is known not to be
conserved in the pure SCF case, and can be taken care of by
recentering the particles.  In this case, every expansion of the SCF
system is taken about its centre of mass, alleviating the need to
recentre the particle.   We check the accuracy of this
modification by monitoring the conservation of linear momentum of the
SCF centre of mass of a spherical stellar system, and the separation
of a binary system of spherical galaxies (one SCF; one Tree) in a
circular and stable orbit.  Figure \ref{binary} shows the separation
between an SCF and a Tree spheroid of equal mass, both with 2500
particles, in a circular orbit.

We show an example of the evolution of the total angular momentum of
the system in Figure (\ref{angmom}a) and although $|{\bf L}|$ is
intrinsically near zero we see that it varies no more than 1\% over
the length of the run.  Figure (\ref{angmom}b) shows the absolute
variation of the $z$-component of ${\bf L}$.  The system in this case
is a Lowered Evans Model with 20000 particles with 10\% Tree
particles.

%\subsection{Linear Momentum}
%\begin{figure}
%\psfig{Figure= linmom.ps,width=8cm,angle=270}
%\caption{Linear momentum figure}
%\label{linmom}
%\end{Figure}

%\subsection{Dynamical Tests}

\subsection{Collapse of uniform sphere}
\label{collapse}
We investigate the performance of the SCFTREE code on a system which
is not initially in equilibrium.  For this purpose we perform
simulations on the collapse of a uniform density spherical
distribution of particles with random velocities scaled such that the
initial virial ratio of the system $\left|2T/W\right|_0=1/2$.
Following the thorough investigation by Hozumi and Hernquist (1995) of
the pure SCF code in similar non-equilibrium states, we trace the
evolution of the virial ratio from its initial value of $1/2$.  Our
purpose here is not to perform a detailed investigation of the
accuracy of the dynamical evolution but as a qualitative check that
the combined SCF and Tree codes behave as expected, and the virial
ratio oscillates about the equilibrium value of
$\left|2T/W\right|_0=1.0$.  The results are shown for a typical run
containing 20000 particles, 10\% of which are randomly allocated as
tree particles.  The truncation parameters for the SCF part are $n=16$
and $l=6$.

\begin{figure}
\psfig{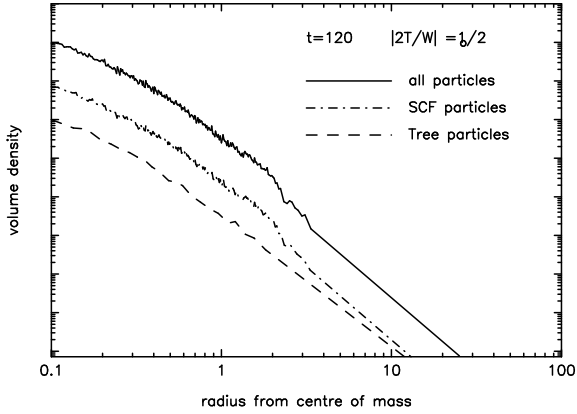}
\caption{Final density profile after $t=120$ of the collapsed sub-virial
($|2T/W|_0=\frac{1}{2}$) uniform density sphere.  Density profiles are
individually shown for the SCF and Tree components, shifted vertically
for comparison.}
\label{density}
\end{figure}

Figure(\ref{density}) shows the plot of final density profile of the
same system after a period of $t=120$.  The separate profiles of the
Tree and SCF systems of particles are shown, together with the total
profile of all of the particles.  In the example shown the system undergoes
homologous collapse (Fillmore \& Goldreich 1984; Gunn 1977) evolving
to a density profile of $\rho \sim r^{-2.5}$.  Towards the core of the
system in this example the number of particles is too small to
adequately resolve the detailed evolution ({\em cf. }  Hozumi \&
Hernquist (1995) who used hundreds of thousands of particles and obtained
adequate resolution  to resolve the flattening of the core.)

\begin{figure}
\psfig{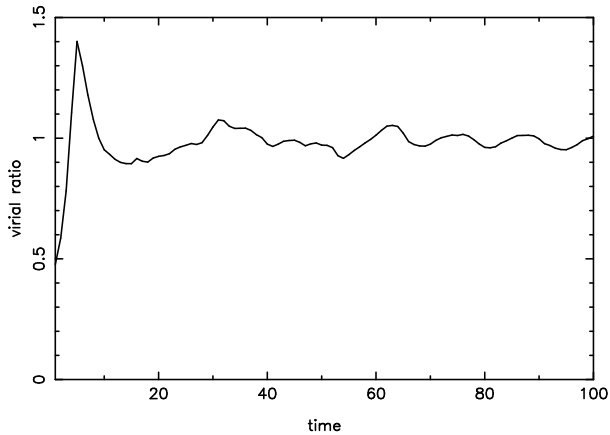}
\caption
{The viral ratio, $|2T/W|$, evolution of the collapse of a subvirial
uniform density sphere.  The total number of particles is 20000, with
a random 10\% being assigned to the Tree system.}
\label{virial}
\end{figure}

\begin{figure}
\psfig{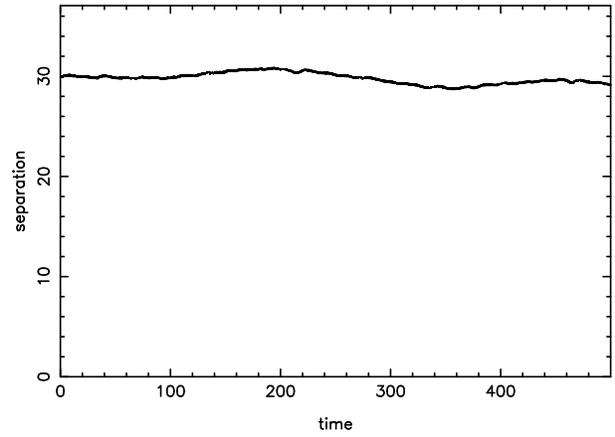}
\caption
{This plot shows the absolute separation between two non-penetrating
systems in circular orbit around each other. Each system is an
identical Lowered Evans model populated with 2500 particles. One is
assigned to the SCF system, the other to the Tree system.  This figure
demonstrates that the method of expanding the SCF potential around its
centre of mass every timestep remains accurate to within 5\% over 400 time
units.}
\label{binary}
\end{figure}

\subsection{Disk + Halo + Bulge}
\label{disk1}
We have seen from the previous tests that the code is efficient and
accurate even when the tree particles are random distributed amongst
the SCF particles.  This is the worst case scenario in terms of
efficiency.  During the force calculation on an SCF particle it will
form a large list of tree cell interactions.  Now we move to a case
where the tree particles form some distinct structure within the SCF
system, that of a disk embedded within a halo.  The study of disk
evolution in a non-interacting system has been the focus of much
research and requires a good deal more attention than a mere
subsection to investigate it fully.  Here we content ourselves with an
example of how the varying of $N_{\rm halo}$ affects the stability of
the disk.  We construct combined disk--halo--bulge models using the
method of Dubinski \& Kuijken (1995).  The ratio of masses of the
components was disk:bulge:halo = 1.00:0.37:12.80, the scale radius of
the disc was set to $R_d=1.0$.  The number of disk particles in each
case was 6000, and the number of bulge particles was 2000.  In four
runs the number of halo particles was varied from 13000 to $10^5$.
Each time the disk particles were assigned to the Tree system and the
rest to the SCF system.  
	
\begin{figure}
\psfig{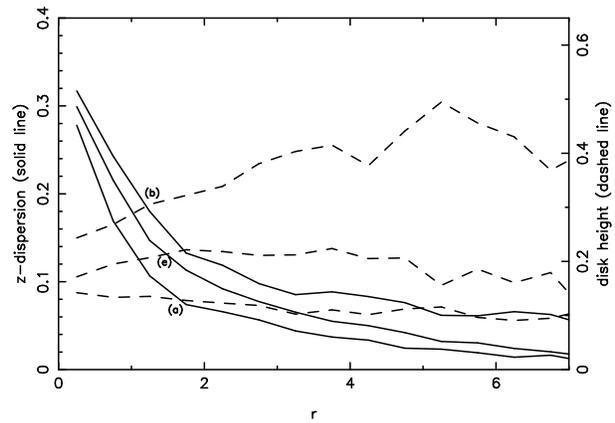}
\caption
{Vertical velocity dispersion, $\sigma_z$ and rms disk height, $z_{\rm
rms}$ as a function of radial distance at $t=120$.  The construction
of the disk--halo--bulge models is described in \S\ref{disk1}.  The
three different runs correspond to the curves (a), (e), and (b) in
Figure \ref{spiral}.}
\label{zdisp}
\end{figure}

\begin{figure*}
\psfig{figure= 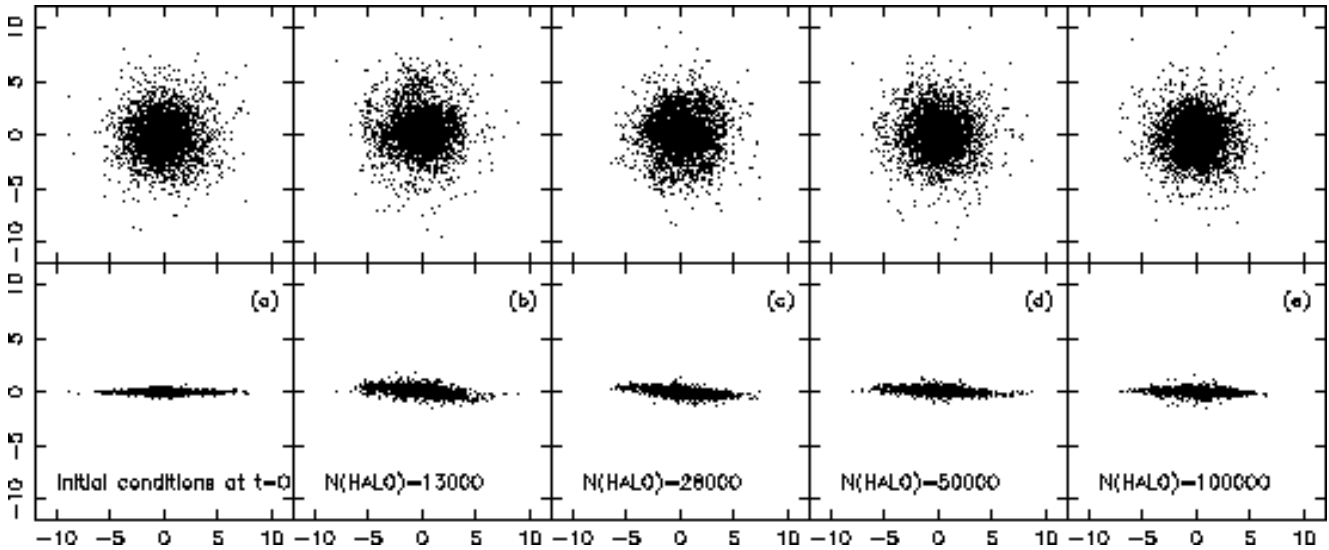,width=17.5cm,angle=270}
\caption{This figure plots the Tree particles only after $t=120$ to
show the effect of changing the number of particles in the halo \ie
the SCF particles, on the stability of the disk. The construction of
the models is described in \S\ref{disk1}.  The initial conditions are
shown in plot (a), with the numbers of particles in the Halo
increasing to the right. The top row and bottom row show the x-y plane
and the y-z plane respectively.}
\label{spiral}
\end{figure*}

Figure \ref{spiral} demonstrates increased stability as number of halo
particles, $N_{\rm halo}$ is increased.  In particular at $N_{\rm
halo}=100000$ little structure has formed in the disk and is unchanged
apart from a slight thickening, whereas for $N_{\rm halo}=13000$ the
disk is particularly unstable to warping, and spiral structure seems
to be developing.  The thickening of the disk shown in Figure
\ref{zdisp}.  These figures confirm what we would expect as we change
the number of $N_{\rm halo}$ in a self-consistent simulation.  With
fewer SCF particles noise in the SCF expanded potential will cause
fluctuations in the halo potential to develop, and thus cause the disk
to become unstable to warping.

\subsection{Dynamical Friction} 
\label{sinking}

\begin{figure}
\psfig{figure= 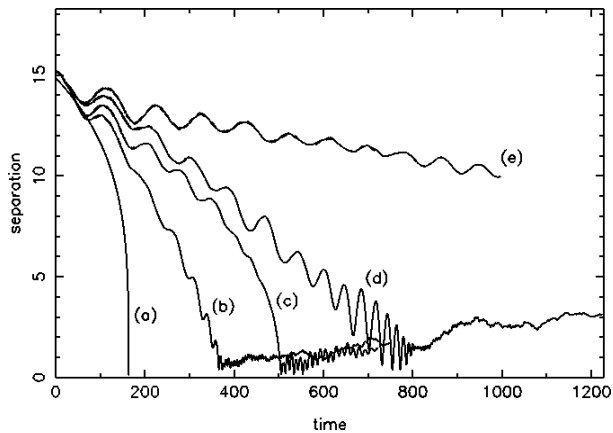,width=8cm,angle=270}
\caption
{The decay in separation between a massive spherical primary, treated
by the SCF system, and a smaller spheroidal satellite, treated by the
Tree system.  The models used for each curve are summarised in Table
1. As discussed in \S\ref{sinking} this figure demonstrates the effectiveness
of using SCFTREE to model dynamical friction.}
\label{sink}
\end{figure}

Penetrating encounters between stellar systems provide many
interesting scenarios which one could fruitfully model with our hybrid
code.  Representing interacting systems separately with the SCF and
Tree methods invites the question of how well SCFTREE describes the
behaviour of SCF and Tree particles interacting with each other in
differing ways.  We address this issue by investigating the ability of
the code to tackle dynamical friction, specifically by running a few
simulations of the sinking satellite sort.  SCF is used to model the
primary, and Tree used for the satellite.  Although the force on each
SCF particle due to the Tree particles is directly calculated (subject
to the tolerance criterion), the Tree particles only respond to the
SCF particles indirectly via the expanded field.  One might initially
be concerned that the lack of direct particle--particle interactions
which account for the deceleration of the satellite will invalidate
the use of this code in such cases, however we show that so long as
the SCF expansion is truncated after a reasonable number of terms
dynamical friction is acceptably treated.

\begin{table}
\begin{center}
\begin{tabular}{clcll}
\hline
Model & $N_{\rm primary}$ & $N_{\rm satellite}$ & \multicolumn{2}{l}{SCF truncation} \\ \hline
(a) & \multicolumn{4}{l}{Approximate analytical solution valid for large r} \\ 
(b) & 20000 & 1000 & $n=16$ & $l=6$ \\
(c) & 20000 & 1 & $n=16$ & $l=6$ \\
(d) & 20000 & 1000 & $n=8$ & $l=2$ \\
(e) & 20000 & 1 & $n=8$ & $l=2$ \\ \hline \hline
Primary: & \multicolumn{4}{l}{Hernquist model; $R_a=1$; $R_{\rm max}=200$; $M=7.39$} \\
Satellite: & \multicolumn{4}{l}{LE model; $R_a=0.1$; $R_{\rm max}=3.6$; $M=0.75$} \\ 
\multicolumn{5}{l}{Initial separation on circular orbits =15} \\
\hline 
\end{tabular}
\caption{Summary of the models and model parameters used in the tests of dynamical friction}
\label{modtbl}
\end{center}
\end{table}

Figure(\ref{sink}) demonstrates the dependence of dynamical
friction on the number of expansion coefficients used in the expanded
SCF potential.  The initial conditions and the models used to
represent the primary and satellite are summarised in
Table {\ref{modtbl}.  Curve (a) in the Figure is a simple analytical
approximation of the sinking rate based on the Chandrasekhar dynamical
friction formula \cite{bt}.  We approximate $f(0)=$constant, and assume
instantaneously circular orbits for the satellite.  This calculation
carries with it the assumption of a fixed primary and is valid only in
the region of large separations, $r$, low circular velocity, no
satellite mass loss, and is to be disregarded at small $r$.

As one might expect when the SCF expansion is truncated at low orders
[curves (d) and (e)] dynamical friction is poorly modelled.
Increasing the truncation order [curves (b) and (c)] with a
corresponding increase in the resolution of the SCF system causes the
Tree system to respond realistically to perturbations in the SCF
density field.  With the SCF expansion truncated at $n=16$ and $l=6$
and taking a single point mass for the satellite (\ie one tree
particle) dynamical friction is modelled well (see curve [c]). We find
these sinking times are consistent with the results obtained by
Hernquist and Weinberg (1989) for fully self-consistent simulations
using a pure tree code and concur with their findings [and of White
(1983)] that when the response of the primary is included in the
numerical calculations the sinking time increases by a factor of
around two.

Decay of the orbit of a single point mass means in simple terms that
orbital energy is transferred to dynamical heating of the particles in
the primary.  For an extended satellite (\ie one composed of
$N_{satellite}$ bodies), orbital energy can also be transferred to the
heating of the satellite, thereby increasing the decay rate and
shortening the decay period.  This interesting result is
demonstrated in curve (b), and certainly warrants greater detailed
exploration.  The result has been clearly alluded to in Weinberg
(1989) in terms of coupling in the analytic linear theory which he
describes. The extended satellite has many weak resonances which can
couple at smaller radii, which the point mass does not.  
	
Finally we note that the sinking satellite loses much of its mass
during its descent.  For the example traced by curve (b) in
Figure(\ref{sink}), 40\% of the Tree particles become unbound from
tree system at $r\sim 7$, and 80\% are unbound at $r\sim 2$.  The
peculiar behaviour of the curves once having reached $r\sim 1$ is due
to the practical disruption of the satellite.

%\section{Application}
%\label{apply}
%Include a science application if we find an easy one.

%Tidal tails to probe dark matter halos, \cite{dm95}
%Maybe an inclined disk in oblate halo \cite{dk95}.

\section{Discussion}
\label{discussion}

In this paper we have presented {\tt SCFTREE}, a hybrid N-body code
combining the Hernquist \& Ostriker Self--Consistent Field code, and
the Barnes--Hut hierarchical tree algorithm.  The SCF code is designed
to model systems with structure resembling to first order the
Hernquist density profile, \mbox{$\rho(r)\propto \frac{1}{r(r+a)^3}$}.
The treecode technique is a proven approach to effectively model systems that
have no {\em a priori} and in effect, no limit to dynamical or spatial
range.  The principle behind the {\tt SCFTREE} scheme is to model
evolving structures that interact with or are within an approximate
Hernquist potential.  The uses of such an approach are wide
ranging, especially modelling the evolution of structures within a
dark matter halo, \eg disks, bars, satellite galaxies, interactions
and encounters, and cluster galaxies within a cluster halo.

The prime advantage of this new technique is a significant improvement
in performance over using pure treecodes which scale as $\cal
O(N\log N)$.  To date only treecode techniques have been adequately
suited to systems with such large dynamical ranges.  However, so much
CPU time is expended on parts of the system with little dynamical
evolution such as the halo.  By representing the halo with the SCF
technique, where the CPU time scales as $\cal O(N)$, we are able to
reduce the overall CPU time and thus can increase the total number of
particles.

As this implementation of the code is based around global geometries
which behave approximately as $R^{1/4}$ in projected density, its use
would not be appropriate to model objects with substantially different
density profiles.  As mentioned in \S \ref{SCF} other basis expansions
for different mass distributions are possible.
However, there are many systems that may be modelled by appropriate
application of {\tt SCFTREE}.

This implementation of {\tt SCFTREE} is able to accurately model
effects of dynamical friction and thus the application to systems of
sinking satellites is a prime application.  Other eminently suitable
applications will be the evolution of inclined disks in flattened
halos \cite{dk95}, the evolution of unstable high-surface-density
disks (Dalcanton \etal 1996; Vine \& Sigurdsson 1997, {\em in
preparation}), and weak encounters between massive elliptical systems
and smaller disk systems (Vine 1997, {\em in preparation}).

This preliminary version of the code has much potential for future
development and expansion, especially in terms of performance.  The
nature of the SCF code makes it suitable for parallelisation \cite{hs95}
together with a vectorised version of the Tree code utilising
heterogeneous systems.  With the advent of parallelised Tree
implementations a doubly parallel SCFTREE is a possibility.

An interesting development will be to experiment with multiple
expansions of the SCF systems.  With such a tool it will be feasible
to model interactions between two or more large spheroidal systems.
For example in the evolution of galaxy groups and multiple mergers
\cite{wh96}.  The principle of having more than one expansion centre
has been exploited some time ago \cite{vv77}.

Work currently in progress includes a powerful improvement to the
algorithm presented here, that of swapping particles between the SCF
and Tree system when one system becomes more appropriate then the
other.  For instance the disruption of a satellite composed of Tree
particles could benefit from transfer of escapers to the main body of
the SCF primary.  Conversely, structure formation in $N>10^8$
cosmological simulations can have increased resolution by assigning
increasing numbers of Tree particles to structures.

Another relatively simple development is to incorporate hydrodynamics
into the Tree part of the code \cite{hk89}.  This could be further
extended to encompass star formation prescriptions
\cite{mh94}. Ultimately it is not unreasonable that the technique
could be employed in cosmological simulations which are now capable of
dealing with much more sophisticated physical processes \cite{kw96}

%\begin{equation}
%\sigma_{\rm ex} = k_{\rm ex}(q_1,q_2) \pi d^2 \cdot { G M_1 F(q_1,q_2) \over
% d} \cdot { 1 \over V_\infty^2 }
%\end{equation}

%\noindent where $q_1=M_2/M_1$, $q_2=M_3/M_1$, and
%$F(q_1,q_2) = q_1(1+q_1+q_2)/q_2(1+q_1)$.  The constant $k_{\rm ex}(q_1,q_2)$
%has to be determined through numerical simulations.

\section*{ACKNOWLEDGEMENTS}
It is a pleasure to thank Lars Hernquist, Chris Mihos, Melvyn Davies,
Ian Bonnell, and Martin Weinberg for their clarifying remarks and
useful discussions.  We are grateful both to Lars Hernquist for making
available to us the original versions of the tree and SCF codes, and
to Konrad Kuijken for the initial conditions code for the Lowered
Evans models.  SGV acknowledges PPARC, the Institute of Astronomy, and
the University of Cambridge for funding during this research, and SS
acknowledges funding from the EU Marie Curie fellowship.

%\appendix

%\section[]{Why put this in at all}
%Hello...

\bsp

\label{lastpage}

\end{document}